\documentclass[12pt,reqno]{article}

\global\arraycolsep=1pt
\usepackage{amsmath,amsfonts}
\usepackage{array}
\usepackage{graphicx}
\usepackage{multicol}
\usepackage[english]{babel}

\usepackage{mathrsfs}
\usepackage[Symbol]{upgreek}
\usepackage{bbm}
\usepackage{dsfont}
\usepackage{amssymb}
\usepackage{wasysym}

\newcommand{\be}{\begin{equation}} 
\newcommand{\ee}{\end{equation}} 
\newcommand{\bea}{\begin{eqnarray}} 
\newcommand{\eea}{\end{eqnarray}}

\newcommand{\bm}[1]{\mbox{\boldmath{$#1$}}}

\def\ie{{\it i.e.\,}}
\def\eg{{\it e.g.\,}}

\def\via{{\it via}\ }

\def\nl{\newline}
\def\ft#1#2{\tfrac{#1}{#2}}

\def\i{{\rm i}}

\newfont{\bbbold}{msbm10}

\def\bbC{\mbox{\bbbold C}}

\def\bbR{\mbox{\bbbold R}}

\def\C{{\mathscr{C}}}
\def\N{{\mathcal{N}}}
\def\M{{\mathcal{M}}}
\def\V{{\mathcal{V}}}
\def\w{{\scriptstyle W}}

\def\DJo{$\;$\kern-.4em \hbox{D\kern-.8em\raise.15ex\hbox{--}\kern.35em okovi\'c}\ }

\def\h{{\boldsymbol{h}}}

\newcommand{\Scal}[1]{\Bigl ({#1} \Bigr )}
\newcommand{\scal}[1]{\bigl ({#1} \bigr )}
\newcommand{\trace}{\hbox {Tr}~}

\newcommand{\imperial}{\it The Blackett Laboratory, Imperial College London\\
Prince Consort Road, London SW7 2AZ}

\newcommand{\auth}{
K.S. Stelle\footnote{email: k.stelle@imperial.ac.uk}}

\begin{document}
\thispagestyle{empty}
\begin{flushright}
\hfill{Imperial/TP/14/KSS/01}\\
\end{flushright} 
\medskip

\begin{center}  

{\Large {\bf Symmetry Orbits of Supergravity Black Holes}} \\
\medskip
In Honor of Andrei Slavnov's $75^{\mbox{\scriptsize th}}$ Birthday

\vspace{15pt}

\auth

\vspace{7pt}
{\imperial} 

\vspace{30pt}

\end{center}  
%%%%%%%%%%%%%%%%%%%
\abstract{Black hole solutions of supergravity theories form families that realizing the deep nonlinear ``duality'' symmetries of these theories. They form orbits under the action of these symmetry groups, with extremal (\ie BPS) solutions at the limits of such orbits. An important technique in the analysis of such solution families employs timelike dimensional reduction and exchanges the stationary black-hole problem for a nonlinear sigma-model problem. Families of extremal or BPS solutions are characterized by nilpotent orbits under the duality symmetries, based upon a tri-graded or penta-graded decomposition of the corresponding duality-group algebra.}
%%%%%%%%%%%%%%%%%%%
\begin{center}
\rule{50pt}{1pt}
\end{center}
\vspace{.3cm}

\section*{\large Andrei Slavnov and the key role of symmetry in physics}

It is a great honor for me to add my congratulations to Professor Andrei Alekseevich Slavnov on the occasion of his $75^{\mbox{\scriptsize th}}$ birthday. As well as being a great friend for most of my scientific life, Andrei's work has shown the pathways in many of my own scientific efforts, on topics ranging from the general theory of gauge fields \cite{Faddeev:1980be} and the crucial importance of symmetry in the renormalization of field theories \cite{Slavnov:1972fg} to the incorporation of higher-derivative terms in quantum field theories \cite{Bakeyev:1996is}, the possible r\^oles of higher dimensions \cite{Slavnov:1995ia,Slavnov:2006rf}, to the relation between unitarity and BRST quantization methods \cite{Slavnov:1989jh}, and to many other central issues in modern theoretical physics. And through all these developments, it has always been a delight to discuss with Andrei, and to share in his great friendliness and warm hospitality during my visits to Russia.

It is accordingly a great pleasure for me to devote this article to Andrei on the occasion of his $75^{\mbox{\scriptsize th}}$ birthday. I look forward to many more years of scientific enlightenment from Andrei, and to many more opportunities to share in his camaraderie.\newpage

\section*{\large Black holes and duality symmetries in supergravity}

Aside from the general mathematical interest in classifying black hole solutions of any kind, the study of families of such solutions is of particular current interest because it touches other important issues in theoretical physics. For example, the classification of BPS and non-BPS black holes forms part of a more general study of branes in supergravity and superstring theory. Branes and their intersections, as well as their worldvolume modes and attached string modes, are key elements in phenomenological approaches to the marriage of string theory with particle physics phenomenology. The related study of nonsingular and horizon-free BPS gravitational solitons is also central to the ``fuzzball'' proposal of BPS solutions as candidate black-hole quantum microstates. 

The search for supergravity solutions with assumed Killing symmetries can be recast as a Kaluza-Klein problem \cite{neukr,Maison,gal}. To see this, consider a 4D theory with a nonlinear bosonic symmetry $G_4$ (\eg the ``duality'' symmetry $\textrm{E}_7$ of maximal $N=8$ supergravity). Scalar fields take their values in a target space $\Phi_4=G_4/H_4$, where $H_4$ is the corresponding linearly realized subgroup, which is generally the maximal compact subgroup of $G_4$ (\eg $\textrm{SU}(8)\subset \textrm{E}_7$ for $N=8$ SG). This search will be constrained by the following considerations:\nl
$\bullet$ We assume that a solution spacetime is asymptotically flat or asymptotically Taub-NUT and that there is a `radial' function $r$ which is divergent in the asymptotic region, $g^{\mu\nu}\partial_\mu r\partial_\nu r \sim 1 + {\cal O}(r^{-1})$.\nl
$\bullet$ Searching for stationary solutions amounts to assuming that a solution possesses a timelike Killing vector field $\kappa_\mu(x)$. Lie derivatives with respect to $\kappa_\mu$ are assumed to vanish on all fields. The Killing vector $\kappa_\mu$ will be assumed to have $W:=-g_{\mu\nu}\kappa^\mu\kappa^\nu\sim 1 + {\cal O}(r^{-1})$.\nl
$\bullet$ We also assume asymptotic hypersurface orthogonality, \ie $\kappa^\nu(\partial_\mu\kappa_\nu-\partial_\nu\kappa_\mu)\sim{\cal O}(r^{-2})$. In any vielbein frame, the curvature will then fall off as $R_{abcd}\sim{\cal O}(r^{-3})$.

The 3D theory obtained after dimensional reduction with respect to a timelike Killing vector $\kappa_\mu$ will have an Abelian principal bundle structure, with a metric
\be
ds^2=-W(dt+B_idx^i)^2 + W^{-1}\gamma_{ij}dx^idx^j
\ee
where $t$ is a coordinate adapted to the timelike Killing vector $\kappa_\mu$ and $\gamma_{ij}$ is the metric on the 3-dimensional hypersurface ${\cal M}_3$ at constant $t$. If the 4D theory also has Abelian vector fields ${\cal A}_\mu$, they similarly reduce to 3D as
\be
4\sqrt{4\pi G}{\cal A}_\mu dx^\mu= U(dt + B_idx^i)+A_idx^i
\ee

The timelike reduced 3D theory will have a $G/H^*$ coset-space structure similar to the $G/H$ coset-space structure of a 3D theory reduced with respect to a  spacelike Killing vector. Thus, for the spacelike reduction of maximal supergravity down to 3D, one obtains an $\textrm{E}_8/\textrm{SO}(16)$ theory from the sequence of dimensional reductions descending from $D=11$ \cite{CJ}.
The resulting 3D theory has this exceptional symmetry because  3D Abelian vector fields can be dualized to scalars; this also happens for the analogous theory subjected to a timelike reduction to 3D. The resulting 3D theory contains 3D gravity coupled to a $G/H^*$ nonlinear sigma model.

Although the numerator group $G$ for a timelike reduction is the same as that obtained in a spacelike reduction, the divisor group $H^*$ for a timelike reduction is a {\em noncompact} form of the spacelike divisor group $H$ \cite{Maison}.
A consequence of this $H\to H^*$ change and of the dualization of vectors is the appearance of {\em negative-sign} kinetic terms for some 3D scalars.

Consequently, maximal supergravity, after a timelike reduction down to 3D and the subsequent dualization of 29 vectors to scalars, has a bosonic sector containing 3D gravity coupled to a nonlinear sigma model with 128 scalar fields. As a consequence of the timelike dimensional reduction and the vector dualizations, however, the scalars do not all have the same signs for their ``kinetic'' terms:\nl
$\bullet$ There are 72 positive-sign scalars: 70 descending directly from the 4D theory, one emerging from the 4D metric and one more coming from the $D=4\rightarrow D=3$ Kaluza-Klein vector, which is subsequently dualized to a scalar.\nl
$\bullet$ There are 56 negative-sign scalars: 28 descending directly from the time components of the 28 4D vectors, and another 28 emerging from the 3D vectors obtained from spatial components of the 28 4D vectors, becoming then negative-sign scalars after dualization.

The sigma-model structure of this timelike reduced maximal theory is $\textrm{E}_8/\textrm{SO}^\ast(16)$. The $\textrm{SO}^\ast(16)$ divisor group is not an $\textrm{SO}(p,q)$ group defined \via preservation of an indefinite metric. Instead it is constructed by starting from the $\textrm{SO}(16)$ Clifford algebra $\{\Gamma^I,\Gamma^J\}=2\delta^{IJ}$ and then by forming the complex $\textrm{U}(8)$-covariant oscillators $a_i := \ft12(\Gamma_{2i-1}+ \i\Gamma_{2i})$ and $a^i\equiv (a_i)^\dag=\ft12(\Gamma_{2i-1}- \i\Gamma_{2i})$. These satisfy the standard fermi-oscillator annihilation/creation anticommutation relations
\be
\{ a_i , a_j\} = \{a^i , a^j \} = 0 \quad , \qquad
\{a_i , a^j \} = \delta_i\,{}^j
\ee

The 120 $\textrm{SO}^\ast(16)$ generators are then formed from the 64 hermitian $\textrm{U}(8)$ generators $a_i\,{}^j$ plus the $2\times 28=56$ antihermitian combinations of  $a_{ij} \pm a^{ij}$. Under $\textrm{SO}^\ast(16)$, the vector representation and the antichiral spinor are pseudo-real, while the 128-dimensional chiral spinor representation is real. This is the representation under which the 72+56 scalar fields transform in the $\textrm{E}_8/\textrm{SO}^\ast(16)$ sigma model.

The 3D classification of extended supergravity stationary solutions \via timelike reduction thus generalizes the 3D supergravity systems obtained from spacelike reduction \cite{de Wit:1992up}. This also connects with $N=2$ models with coupled vectors  \cite{Meessen:2006tu}
and $N=4$ models with vectors, where solutions have also been generated using duality symmetries \cite{Cvetic:1995uj,Cvetic:1995yq}

\subsection*{\normalsize Stationary solutions and harmonic maps}

The process of timelike dimensional reduction down to 3 dimensions together with dualization of all form-fields to scalars produces an Euclidean gravity theory coupled to a $G/H^\ast$ nonlinear sigma model, $I_\sigma = \int d^3x \sqrt\gamma (R(\gamma) - \ft12 G_{AB}(\phi)\partial_i\phi^A\partial_j\phi^B\gamma^{ij})$, where $G_{AB}(\phi)$ is the $G/H^\ast$ sigma-model target-space metric and \(\gamma_{ij}\) is the 3D metric. Varying this action produces the 3D field equations
\bea
&&{1\over\sqrt\gamma}\nabla_i(\sqrt\gamma\gamma^{ij}G_{AB}(\phi)\partial_j\phi^B) = 0\\
&&R_{ij}(\gamma) = \ft12 G_{AB}(\phi)\partial_i\phi^A\partial_j\phi^B
\eea
where $\nabla_i$ is a doubly covariant derivative (both for the 3D space ${\cal M}_3$ and for the $G/H^\ast$ target space).

Now one can make the simplifying assumption that $\phi^A(x)=\phi^A(\sigma(x))$, depending on a single intermediate map $\sigma(x)$. Subject to this assumption, the field equations  become
\bea
\nabla^2\sigma {d\phi^A\over d\sigma}+\gamma^{ij}\partial_i\sigma\partial_j\sigma[{\partial^2\phi^A\over d\sigma^2}+\Gamma^A_{BC}(G){d\phi^B\over d\sigma}{d\phi^C\over d\sigma}] &=& 0\\
R_{ij} = \left(\ft12 G_{AB}(\phi){d\phi^A\over d\sigma}{d\phi^B\over d\sigma}\right)\partial_i\phi^A\partial_j\phi^B&&
\eea
Now one can use the gravitational Bianchi identity $\nabla^i(R_{ij}-\ft12\gamma_{ij}R)\equiv0$ to obtain
\(
\ft14 {d\over d\sigma}(G_{AB}(\phi){d\phi^A\over d\sigma}{d\phi^B\over d\sigma})(\nabla^i\sigma\partial_i\sigma)=0\ .
\)
Requiring separation of the $\sigma(x)$ properties from the ${d\over d\sigma}$ properties thus leads to the conditions
\bea
\nabla^2\sigma &=& 0\label{first}\\
{d^2\phi^A\over d\sigma^2} + \Gamma^A_{BC}(G){d\phi^B\over d\sigma}{d\phi^C\over d\sigma} &=& 0\label{second}\\
{d\over d\sigma}\left(G_{AB}(\phi){d\phi^A\over d\sigma}{d\phi^B\over d\sigma}\right) &=& 0\label{third}
\eea

The first equation \eqref{first} above  implies that $\sigma(x)$ is a {\em harmonic map} from the 3D space ${\cal M}_3$ into a curve $\phi^A(\sigma)$ in the $G/H^\ast$ target space.
The second equation \eqref{second} implies that $\phi^A(\sigma)$ is a {\em geodesic} in $G/H^\ast$.
The third equation \eqref{third} implies that $\sigma$ is an {\em affine parameter}.
The decomposition of $\phi:  {\cal M}_3\to G/H^\ast$ into a harmonic map $\sigma:{\cal M}_3\to \bbR$ and a geodesic $\phi:\bbR\to G/H^\ast$ is in accordance with a general theorem on harmonic maps \cite{Eells}
according to which the composition of a harmonic map with a totally geodesic one is again harmonic.
Such factorization into geodesic and harmonic maps is also characteristic of general higher-dimensional $p$-brane supergravity solutions \cite{neukr,gal}.

Here is a sketch of the map composition:
\begin{center}
\includegraphics[scale=1.5]{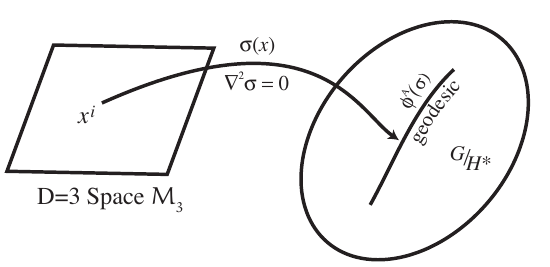}
\end{center}

As an example \cite{Maison} of the way in which the above construction emerges naturally for single-center solutions, consider 4D gravity together with an Abelian $\textrm{U}(1)$ vector field, \ie 4D Maxwell-Einstein theory. One can search for stationary solutions depending on a single intermediate variable. For this intermediate variable, one starts simply with the radius $r=\sqrt{x^ix^i}$, so in supposing that solutions have just this radial dependence, one is actually looking for spherically symmetric solutions with isometry group $\textrm{SO}(3)$. Changing to 3D polar coordinates, the metric on ${\cal M}_3$ can be parametrized as $ds^2=\gamma_{ij}dx^idx^j=dr^2+f(r)^2(d\vartheta^2+\sin^2\vartheta d\varphi^2)$. The reduced 3D equations of motion then become
\bea
f^{-2}{d\over dr}(f^2{d\phi^A\over dr})+\Gamma^A_{BC}(\phi){d\phi^A\over dr}{d\phi^B\over dr} =  0&&\label{eqna}\\
R_{rr}=-2f^{-1}{d^2f\over dr^2} = G_{AB}(\phi){d\phi^A\over dr}{d\phi^B\over dr}&&\label{eqnb}\\
R_{\varphi\varphi}=R_{\theta\theta}=f^{-2}({d\over dr}(f{df\over dr})-1) = 0\ .\label{eqnc}&&
\eea

Equation \eqref{eqnc} has the general solution $f(r)^2=(r-r_0)^2+c^2$.
Introducing then $\sigma(r):=-\int_r^\infty f^{-2}(s)ds$, one obtains a harmonic intermediate function on ${\cal M}_3$ with respect to with the metric $\gamma_{ij}$, and equation \eqref{eqna} then becomes
\be
{d^2\phi^A\over d\sigma^2}+\Gamma^A_{BC}(\phi){d\phi^B\over d\sigma}{d\phi^C\over d\sigma} = 0
\ee
with $\phi^A(r)=\phi^A(\sigma(r))$.
This is the equation for a geodesic in the 4-dimensional symmetric space $G/H^*=\textrm{SU}(2,1)/\textrm{S}(\textrm{U}(1,1)\times\textrm{U}(1))$, with signature $(++--)$. Owing to the indefinite character of this sigma-model target space, there are a variety of different solution orbits, depending on the ``spacelike'', ``lightlike'' or ``timelike'' character of the geodesic $\phi^A(\sigma)$.

Restricting attention to the subspace of static solutions with electric charge only (magnetic charge can also be included by a duality transformation), the relevant sigma-model structure simplifies to $(G/H^*)_{\textrm{static}}=\textrm{SO}(2,1)/\textrm{SO}(1,1)$. The line element in this two-dimensional target space is $ds^2={d\Delta^2\over2\Delta^2}-{2dA^2\over\Delta}$, where $\Delta$ and $A$ are respectively the gravitational and electric potentials. This is actually just the metric for two-dimensional de Sitter space, for which the corresponding geodesic equations are 
\be
\ddot\Delta-\Delta^{-1}\dot\Delta^2-2\dot A^2=0\qquad\qquad \ddot A-\Delta^{-1}\dot\Delta\dot A=0\ ;
\ee
these can be explicitly solved subject to the boundary conditions $\Delta(0)=1$, $A(0)=0$, corresponding to the desired asymptotic behavior  as $r\to\infty$.

In this way, one obtains three families of Reissner-Nordstrom solutions, with solution classes separating according to the sign of the integration constant $v^2=\frac12 G_{AB}{d\phi^A\over d\sigma}{d\phi^B\over d\sigma}$ characterizing the geodesic on $\textrm{SO}(2,1)/\textrm{SO}(1,1)$ as spacelike ($v^2>0$), lightlike ($v^2=0$) or timelike ($v^2<0$).

To understand such solution orbits more generally \cite{Bossard:2009at}, we need to define the appropriate ``charges'' that will characterize the individual solutions. Towards this end, consider the Komar two-form $K \equiv  \partial_\mu  \kappa_\nu dx^\mu \wedge dx^\nu$. This is invariant under the action of the timelike isometry and, by the asymptotic hypersurface orthogonality assumption, is asymptotically horizontal. This condition is equivalent to the requirement that the scalar field $B$ dual to the Kaluza-Klein vector arising out of the 4D metric must vanish like ${\cal O}(r^{-1})$ as $r\to\infty$. In this case, one can define the Komar mass and NUT charge by (where $s^*$ indicates a pull-back to a section) \cite{Bossard:2008sw}
\be 
m \equiv \frac{1}{8 \pi }Ê\int_{\partial {\cal M}_3}  s^* \star K  \hspace{10mm}  n 
\equiv \frac{1}{8 \pi }Ê\int_{\partial {\cal M}_3} s^* K \ .
\ee

The Maxwell field also defines charges. Using the Maxwell field equation $d \star \mathcal{F}  = 0$, where $\mathcal{F }\equiv\delta{\cal{L}}/\delta F$ 
is a linear combination of the two-form field strengths $F$ depending on the 4D scalar fields, and using the Bianchi identity $d F = 0$, one obtains conserved electric and magnetic charges:
\be 
q \equiv  \frac{1}{2 \pi }Ê\int_{\partial {\cal M}_3}  s^* \star \mathcal{F}  
\hspace{10mm} p \equiv  \frac{1}{2 \pi }Ê\int_{\partial {\cal M}_3}  s^* F \ .
\ee

Now consider these charges from the three-dimensional point of view in order to clarify their transformation properties under the 3D duality group $G$. 
The three-dimensional theory is described in terms of a coset representative $\V\in G/ H^*$. The Maurer--Cartan form $\V^{-1} d \V$ for the Lie algebra $\mathfrak{g}$
decomposes as
\be
\V^{-1} d\V = Q + P \quad, \qquad Q\equiv Q_\mu dx^\mu\in \mathfrak{h}^* 
  \;\; , \;\;\; 
 P\equiv P_\mu dx^\mu\in \mathfrak{g}\ominus\mathfrak{h}^* \ .
\ee
Then the three-dimensional scalar-field equation of motion can be rewritten as $d \star \V P  \V^{-1} = 0$, so the $\mathfrak{g}$-valued ``Noether current'' is $\star \V P \V^{-1}$. Since the three-dimensional theory is Euclidean, one cannot really properly speak 
of a conserved charge. Nevertheless, since $\star \V P \V^{-1}$ is $d$-closed, 
the integral of this $2$-form over a given homology cycle does not depend on 
the particular representative of that cycle. So in this sense we may consider that the integral over this 2-form yields a charge.

Accordingly, for stationary solutions, the integral of this 3D 2-form current $\star \V P \V^{-1}$, taken over any spacelike closed surface $\partial {\cal M}_3$ containing 
in its interior all the singularities and topologically non-trivial subspaces of a given
solution, defines a $\mathfrak{g} \ominus \mathfrak{h}^*$-valued Noether-charge matrix $\C$:
\be
\C \equiv \frac{1}{4 \pi} \int_{\partial {\cal M}_3} \star \V P \V^{-1} \ .
\ee
This transforms in the adjoint representation of the duality group $G$ in accordance with the standard non-linear action of $G$ on $\V\in G/ H^*$. For asymptotically-flat solutions, $\V$ can be arranged to tend asymptotically at infinity to the identity matrix. The charge matrix $\C$ in that case is simply given by the asymptotic value 
of the one-form $P$:
\be
P = \C \, \frac{ dr }{r^2} + \mathcal{O}(r^{-2}) \ .
\ee 

\subsection*{\normalsize Graded structure}

Now let us follow the evolution of the duality group $G$ down a couple of steps in dimensional reduction. In $D=5$, maximal supergravity has the maximally noncompact duality group $\textrm{E}_{6,6}$, with the 42 $D=5$ scalar fields taking their values in the coset space $\textrm{E}_{6,6}/\textrm{USp}(8)$, while the 1-form (\ie vector) fields transform in the $\bm{27}$ of $\textrm{E}_{6,6}$.

 Proceeding on down to 4D, the 27 $D=5$ vectors produce new scalars upon dimensional reduction, and one also gets a new Kaluza-Klein scalar emerging from the $D=5$ metric, making up the total of 70 scalars in the 4D theory. These take their values in $\textrm{E}_{7,7}/\textrm{SU}(8)$, while the 4D vector field strengths transform in the $\bm{56}$ of $\textrm{E}_{7,7}$. The new KK scalar corresponds to a $\mathfrak{gl}_1$ {\em grading generator} of $\textrm{E}_{7,7}$, leading to a tri-graded decomposition of the $\textrm{E}_{7,7}$ algebra as follows:
\be
\mathfrak{e}_{7,7}\simeq \overline{\bm{27}}^{(-2)}\oplus(\mathfrak{gl}_1\oplus\mathfrak{e}_{6,6})^{(0)}\oplus \bm{27}^{(2)}
\ee
where the superscripts indicate the $\mathfrak{gl}_1$ grading.

Continuing on down to 3D \via a timelike reduction, one encounters a new phenomenon: 3D vectors can now be dualized to scalars. This is already clear in the timelike reduction of pure 4D GR to 3D, where one obtains a two-scalar system taking values in $\textrm{SL}(2,\bbR)/\textrm{SO}(2)$, where $\textrm{SL}(2,\bbR)$ is the Ehlers group \cite{Ehlers}. Its generators can be written 
\be  \gamma \bm{h}\oplus \epsilon \bm{e}\oplus \varphi \bm{f} = \begin{pmatrix}
\gamma &\epsilon\\ \varphi &-\gamma
\end{pmatrix}
\ee
 and its Lie algebra is \([\bm{h},\bm{e}]=2\bm{e}\ ,\quad [\bm{h},\bm{f}]=-2\bm{f}\ ,\quad [\bm{e},\bm{f}]=\bm{h}\).

 Accordingly, in reducing from 4D to 3D a supergravity theory with 4D symmetry group $G_4$, with corresponding Lie algebra $\mathfrak{g_4}$ and with vectors transforming in the $\mathfrak{l}_4$ representation of $\mathfrak{g_4}$, one obtains a penta-graded structure for the 3D Lie algebra $\mathfrak{g}$, with the Ehlers generator $\bm{h}$ now acting as the grading generator $\bm{1}^{(0)}$:
\be
\mathfrak{g}\simeq \bm{1}^{(-2)}\oplus \overline{\mathfrak{l}_4}^{(-1)}\oplus(\bm{1}\oplus\mathfrak{g}_4)^{(0)}\oplus  \mathfrak{l}_4^{(+1)}\oplus \bm{1}^{(2)}
\ee
For example, in 3D maximal supergravity one obtains in this way $\mathfrak{e}_{8,8}$:
\be
\mathfrak{e}_{8,8}\simeq \bm{1}^{(-2)}\oplus \overline{\bm{56}}^{(-1)}\oplus(\bm{1}\oplus\mathfrak{e}_{7,7})^{(0)}\oplus  \bm{56}^{(+1)}\oplus \bm{1}^{(2)}\quad\hbox{(248 generators)}
\ee

Now apply this to the decomposition of the coset-space structure for the 3D scalar fields and for the charge matrix $\C$. In 4D, the scalars are associated to the coset generators $\mathfrak{g}_4 \ominus \mathfrak{h}_4 $, where $\mathfrak{h}_4$ is the Lie algebra of the 4D divisor group $H_4$. The representation carried by the 4D electric and magnetic charges $q$ and $p$ is $\mathfrak{l}_4$. Then the 3D scalars and the charge matrix $\C$ can be decomposed into three irreducible representations with respect to $\mathfrak{so}(2) \oplus \mathfrak{h}_4$ according to 
\be
\mathfrak{g} \ominus \mathfrak{h}^*
\cong \scal{  \mathfrak{sl}(2,\mathds{R})\ominus \mathfrak{so}(2) }  
\oplus \mathfrak{l}_4 \oplus \scal{Ê\mathfrak{g}_4 \ominus \mathfrak{h}_4 }
\ee 
The metric induced by the $\mathfrak{g}$ algebra's Cartan-Killing metric on this coset space is positive definite for the 
first and last terms, but is negative definite for $\l_4$. One associates the $\mathfrak{sl}(2,\mathds{R}) 
\ominus \mathfrak{so}(2)$ components to the Komar mass and the Komar NUT 
charge, while the $\mathfrak{l}_4$ components are associated to the electromagnetic charges. 
The remaining $\mathfrak{g}_4 \ominus \mathfrak{h}_4$ charges belong to the
Noether current of the 4D theory.

\subsection*{\normalsize Characteristic equation}

Breitenlohner, Gibbons and Maison \cite{Maison} proved that if $G$ is simple, all the non-extremal single-black-hole solutions of a given theory lie on the $H^*$ orbit of a Kerr solution. Moreover, all {\it static} solutions regular outside the horizon with a charge matrix satisfying $\trace \C^2 > 0$ 
lie on the $H^*$-orbit of a Schwarzschild solution. (Turning on and off angular momentum requires consideration of the $D=2$ duality group generalizing the Geroch $A_1{}^1$ group; we shall not go into that symmetry structure here.)

Using Weyl coordinates, where the 4D metric takes the form 
\be
ds^2=f(x,\rho)^{-1}[e^2k(x,\rho)(dx^2+d\rho^2)+\rho^2d\phi^2]+f(x,\rho)(dt+A(x,\rho)d\phi)^2\ ,
\ee
the coset representative $\V$ associated to the 
Schwarzschild solution with mass $m$ can be 
written in terms of the non-compact generator $\h$ of the Ehlers
$\mathfrak{sl}(2,\mathds{R})$ alone, \ie
\be
\V = \exp \left(\frac12 \ln \frac{r-m}{r+m} \, \h \right) 
\quad\to\qquad \C = m \h\ .
\ee

For the maximal $N=8$ theory with symmetry $\textrm{E}_{8(8)}$ (and also for the exceptional `magic'  $N=2$ supergravity 
\cite{Gunaydin:1983rk} with symmetry $\textrm{E}_{8(-24)}$), one has $h=\mathrm{diag}[2,1,0,-1,-2]$, so
\vskip-.4cm
\be 
\h^5 = 5 \h^3 - 4 \h 
\ee
Consequently, the charge matrix $\C$  satisfies in all cases the characteristic equation
\be 
\quad \C^5 = 5 c^2 \C^3 - 4 c^4  \C
\ee
where $c^2 \equiv \frac{1}{\phantom{|}\scriptsize{\trace}(\h^2)} \, \trace(\C^2)$ is the {\it extremality parameter} ($c^2=0$ for extremal static solutions, while $c^2=m^2$ for Schwarzschild).
Moreover, for all but the two exceptional $\textrm{E}_8$ cases, a stronger constraint is actually satisfied by the charge matrix $\C$:
\be
\label{cubic} \quad 
\C^3 = c^2 \C\ .
\ee
The characteristic equation selects acceptable orbits of solutions, \ie orbits not exclusively containing solutions with naked singularities. It determines $\C$ in terms of the mass and NUT charge and the 4D electromagnetic charges.

The parameter $c^2$ is the same as the $\hbox{(target space velocity)}^2$ of the above harmonic-map discussion: $c^2=v^2$. The Maxwell-Einstein theory is the simplest example with an indefinite-signature sigma-model metric, with a scalar-field target space $G/H^\ast=\textrm{SU}(2,1)/ \textrm{S(U}(1,1)\times \textrm{U}(1))$. The Maxwell-Einstein charge matrix is
\be
\C_{\hbox{\scriptsize ME}}=\begin{pmatrix} m & n & -z/\sqrt{2} \\ n & -m & \i z/\sqrt{2} \\
\bar z/\sqrt{2} & \i \bar z/\sqrt{2} & 0 
\end{pmatrix} \in su(2,2)\ominus u(1,1)
\ee
where $z=q+\i p$ is the complex electromagnetic charge. The Maxwell-Einstein extremality parameter is $c^2=m^2+n^2-z\bar z$. Solutions fall into three categories: $c^2>0$ nonextremal, $c^2=0$ extremal and $c^2<0$ hyperextremal. The hyperextremal solutions have naked singularities, while the nonextremal and extremal solutions have their singularities cloaked by horizons.

\subsection*{\normalsize Dirac equation}

Extremal solutions have $c^2=0$, implying that the charge matrix $\C$ becomes {\em nilpotent}: $\C^5 = 0$ in the $\textrm{E}_8$ cases and 
$\C^3 = 0$ otherwise.

For $\N$ extended supergravity theories, one finds $H^*\cong \textrm{Spin}^*(2\N)\times H_0$ and the charge matrix $\C$ transforms as a Weyl spinor of $\textrm{Spin}^*(2\N)$, also valued in a representation of $\mathfrak{h}_0$ (where $\mathfrak{h}_0$ acts on the matter content of reducible $N=4$ theories). As in the $\textrm{SO}^\ast(16)$ case considered earlier, one defines the $\textrm{Spin}^*(2\N)$ fermionic oscillators
\be
a_i := \frac12 \Big(\Gamma_{2i-1} + \i\Gamma_{2i}\Big)
 \qquad 
a^i \equiv (a_i)^\dagger = \frac12\Big(\Gamma_{2i-1} - \i\Gamma_{2i}\Big)
\ee
for $i,j,\dots = 1,\dots , \N$. These obey standard fermionic annihilation \& creation anticommutation relations. Using this annihilation/creation oscillator basis, the charge matrix $\C$ can be represented as a state (where $a_i\left| 0\right>=0$)
\be 
\left| \C \right> \equiv \Scal{Ê\w + Z_{ij} a^i a^j + 
\Sigma_{ijkl} a^i a^j a^k a^l + \cdots }Ê \left| 0 \right> \ .
\ee

From the requirement that the dilatino fields be invariant under the unbroken supersymmetry of a BPS solution, one derives a `Dirac equation' for the charge state vector,
\be
\Big( \epsilon^i_\alpha a_i + 
\Omega_{\alpha\beta} \epsilon_i^\beta a^i \Big) |\C\rangle = 0  
\ee 
where $(\epsilon^i_\alpha,\epsilon_i^\alpha)$ is the asymptotic 
(for $r\to\infty$) value of the Killing spinor and where $\Omega_{\alpha\beta}$ is a symplectic form on $\bbC^{2n}$ in cases with $n/N$ preserved supersymmetry. 
This condition turns out to be equivalent to the algebraic requirement that $\C$ be a {\em pure spinor} of  $\textrm{Spin}^*(2\N)$. For BPS solutions, it has the consequence that the characteristic equations can be explicitly solved in terms of rational functions.

Note that $c^2=0 \Longleftrightarrow \left<\C|\C\right > =0$ is a {\em weaker} condition than the supersymmetry Dirac equation. Extremal and BPS are not always synonymous conditions, although they coincide for $\N\le 5$ pure supergravities. They are not synonymous, for example, for $\N=6\, \& \,8$ or for theories with vector matter coupling.

\subsection*{\normalsize BPS Strata}

Analysis of the `Dirac equation' or the nilpotency degree of the charge matrix $\C$  leads to a decomposition of the moduli space $\M$ of supergravity solutions into {\em strata} of various BPS degrees. Letting $\M_0$ be the non-BPS stratum, $\M_1$ being the $\ft1N$ BPS stratum, etc., the dimensions of some of the strata for pure supergravity theories turn out to be \cite{Bossard:2009at}
 
\begin{gather}
\begin{array}{|c|c|c|c|c|c|c|}
\hline
  & \N=2 & \N=3  & \N=4 \,\, & \N=5 & \N=6 &\N=8 \\*
\hline
& & & & & & \vspace{-4mm} \\*
{\rm dim} ( \mathcal{M}_0) Ê \hspace{1.5mm} &\hspace{3mm} 4  \hspace{3mm} & \hspace{1.5mm} 8  \hspace{1.5mm}  & \hspace{1.5mm}  14  \hspace{1.5mm}  & \hspace{1.5mm} 22  \hspace{1.5mm}  & \hspace{1.5mm} 34  \hspace{1.5mm}   & \hspace{1.5mm} 58  \hspace{1.5mm}   \\*
& & & & & & \vspace{-4mm} \\*
 \hline
& & & & & &\vspace{-4mm} \\*
{\rm dim} ( \mathcal{M}_1) Ê \hspace{1.5mm} &\hspace{3mm} 3  \hspace{3mm} & \hspace{1.5mm} 7  \hspace{1.5mm}  & \hspace{1.5mm}  13  \hspace{1.5mm}  & \hspace{1.5mm} 21  \hspace{1.5mm} & \hspace{1.5mm} 33  \hspace{1.5mm}  & \hspace{1.5mm} 57  \hspace{1.5mm}   \\*
& & & & & & \vspace{-4mm} \\*
\hline
& & & &  & &\vspace{-4mm} \\*
{\rm dim} ( \mathcal{M}_1^0) Ê \hspace{1.5mm} &\hspace{3mm}      \hspace{3mm} & \hspace{1.5mm}    \hspace{1.5mm}  & \hspace{1.5mm}   \hspace{1.5mm}  & \hspace{1.5mm}    \hspace{1.5mm} & \hspace{1.5mm} 32  \hspace{1.5mm} & \hspace{1.5mm} 56   \hspace{1.5mm} \\*
& & & & & & \vspace{-4mm} \\*
 \hline
& & & &  & &\vspace{-4mm} \\*
{\rm dim} ( \mathcal{M}_2) Ê \hspace{1.5mm} &\hspace{3mm}      \hspace{3mm} & \hspace{1.5mm}    \hspace{1.5mm}  & \hspace{1.5mm}  8  \hspace{1.5mm}  & \hspace{1.5mm} 16   \hspace{1.5mm} & \hspace{1.5mm} 26  \hspace{1.5mm} & \hspace{1.5mm} 46   \hspace{1.5mm} \\*
& & & & & & \vspace{-4mm} \\*
 \hline
& & & &  & &\vspace{-4mm} \\*
{\rm dim} ( \mathcal{M}_4) Ê \hspace{1.5mm} &\hspace{3mm}      \hspace{3mm} & \hspace{1.5mm}    \hspace{1.5mm}  & \hspace{1.5mm}     \hspace{1.5mm}  & \hspace{1.5mm}     \hspace{1.5mm} & \hspace{1.5mm} 17  \hspace{1.5mm} & \hspace{1.5mm} 29   \hspace{1.5mm} \vspace{-4mm} \\*
& & & &  & & \\*
 \hline
\end{array} \nonumber 
\end{gather}

Where do such stratum dimensions come from? Take the non-extremal stratum of $N=8$ supergravity as an example, with 58 moduli. In order for this small number to be related to an $\textrm{E}_8$ group action, one needs to find a {\em large} isotropy group to divide by.  The existence of such a large subgroup is a peculiarity of non-compact groups, analogous to the 4-generator Borel subgroup of the 4D Lorentz group. For the non-extremal $N=8$ supergravity stratum ${\cal M}_0$, there is a 190 generator parabolic subgroup ${\cal P}_0$ containing the 4D duality group $\textrm{E}_7$, 56 generators corresponding to the 56 electromagnetic charges of the 4D theory, plus one more generator. The resulting $\textrm{E}_8/{\cal P}_0$ coset is then 58 dimensional. However, as we shall see, this gives a proper group action only on a dense subset of the full moduli space. Analysis of the extremal strata of supergravity solutions requires understanding the {\em nilpotent orbits} of the 3D duality group $H^*$. This analysis links up with the established mathematical literature on nilpotent orbits, in particular by \DJo \cite{DJo}.

\subsection*{\normalsize Almost Iwasawa decompositions}

Earlier analysis of the orbits of the 4D symmetry groups $G_4$ \cite{Cremmer:1980zc} heavily used the Iwasawa decomposition 
\vskip-.4cm
\be
g  = u_{(g,Z)}  \, \exp\Big( \ln \lambda_{(g,Z)} \, \boldsymbol{z}  \Big) \, 
b_{(g, Z)}  
\ee
with $u_{(g,Z)}\in H_4$ and $b_{(g,Z)}\in\mathfrak{B}_Z$ where $\mathfrak{B}_Z\subset G_4$ is the parabolic subgroup that leaves the charges
$Z$ invariant up to a multiplicative factor $\lambda_{(g,Z)}$. This multiplicative factor can be compensated for by `trombone' transformations combining Weyl scalings with compensating dilational coordinate transformations, leading to a formulation of active symmetry transformations that map solutions into other solutions with {\em unchanged asymptotic values} of the spacetime metric and scalar fields.

The 4D `trombone' transformation finds a natural home in the parabolic subgroup of the 3D duality group $G$. The 3D structure is characterized by the fact that the Iwasawa decomposition {\em breaks down} for noncompact divisor groups $H^*$. The Iwasawa decomposition does, however work ``almost everywhere'' in the 3D solution space. The places where it fails are precisely the extremal suborbits of the duality group. This has the consequence that $G$ {\it does not act transitively on its own orbits}. There are $G$ transformations which allow one to send $c^2\to 0$, thus landing on an extremal (generally BPS) suborbit. However, one cannot then invert the map and return to a generic non-extremal solution from the extremal solution reached on a given $G$ trajectory.

\subsection*{\normalsize Multi-centered solutions}

The above framework applies equally to multi-centered as to single-centered solutions \cite{Bossard:2009my,Bossard:2011kz}. One may start from a general ansatz
\be 
\V (x)=   \V_0 \, \exp(- \sum_n \mathcal{H}^n (x) \C_n )
\ee
with Lie algebra elements $\C_n \in  \mathfrak{g} \ominus \mathfrak{h}^*$ and functions $\mathcal{H}^n(x)$ to be determined by the equations of motion. Defining as above 
$\V^{-1} d\V = Q + P$ and restricting $P$ to depend linearly on the functions $\mathcal{H}^n(x)$, one finds the requirement
$ [Ê\C_m , [ \C_n , \C_p ] ] = 0$.
The Einstein and scalar equations of motion then reduce to 
\be 
R_{\mu\nu}  - \frac12 g_{\mu\nu} R = \sum_{mn} \partial_\mu \mathcal{H}^m \partial_\nu \mathcal{H}^n \, \hbox{Tr}\  \C_m \C_n  \qquad d \star d \mathcal{H}^n = 0 \ .
\ee
Restricting attention to solutions where the 3-space is flat then requires $\hbox{Tr}\   \C_m \C_n = 0$. The resulting system generalizes that of Reference \cite{gal}. Solving $ [Ê\C_m , [ \C_n , \C_p ] ] = 0=\hbox{Tr}\   \C_m \C_n$ is now reduced to an algebraic problem amenable to the above nilpotent-orbit analysis: multi-centered non-extremal and extremal stationary solutions can accordingly be formed from extremal single-centered constituents.

\subsection*{\normalsize Summary}

What has been developed is a quite general framework for the analysis of stationary supergravity solutions using duality-symmetry orbits.
The Noether-charge matrix $\C$ satisfies a characteristic equation $\C^5 = 5 c^2 \C^3 - 4 c^4  \C$ in the maximal $\textrm{E}_8$ cases and $\C^3 = c^2 \C$ in the non-maximal cases, where $c^2 \equiv \frac{1}{\phantom{|}\scriptsize{\trace} \h^2} \, \trace \C^2$ is the extremality parameter.
Extremal solutions are characterized by $c^2=0$, and $\C$ becomes nilpotent ($\C^5=0$ or $\C^3=0$) on the corresponding extremal suborbits.
BPS solutions have a charge matrix $\C$ satisfying an algebraic `supersymmetry Dirac equation' which encodes the general properties of such solutions. This is a stronger condition than the $c^2=0$ extremality condition.
The orbits of the 3D duality group $G$ are not always acted upon transitively by $G$. This is related to the failure of the Iwasawa decomposition for noncompact divisor groups $H^*$. The Iwasawa failure set corresponds to the extremal suborbits.

\end{document}